# Optimum Bi-level Hierarchical Clustering for Wireless Mobile Tracking Systems

Uthman Baroudi[1], Abdulrahman Abu Elkhail[2], Hesham Alfares[3]*

King Fahd University of Petroleum and Minerals

Dhahran, Saudi Arabia

1: ubaroudi@kfupm.edu.sa, 2: g201536490@kfupm.edu.sa, 3: alfares@kfupm.edu.sa

*Corresponding author

*Abstract*

A novel technique is proposed to optimize energy efficiency for wireless networks based on hierarchical mobile clustering. The new bi-level clustering technique minimizes mutual interference and energy consumption in large-scale tracking systems used in large public gatherings such as festivals and sports events. This technique tracks random movements of a large number people in a bounded area by using a combination of smart-phone Bluetooth and Wi-Fi connections. It can be effectively used for monitoring health conditions of crowd members and providing their locations and movement directions. An integer linear programming (ILP) model of the problem is formulated to optimize the formation of clusters in a two-level hierarchical structure. In order to evaluate the proposed technique, it is compared to the optimum solutions obtained from the ILP model for both single-level and two-level clustering. Moreover, a Matlab/Simulink simulation model is developed and used to test the technique's performance under realistic operating conditions. The results demonstrate a very good performance of the proposed technique.

*Keywords*

Tracking; Wireless sensor networks, Integer programming; Bluetooth and Wi-Fi interference; Hierarchical clustering.

# 1. Introduction

Modern smartphone devices are capable of activating tracking systems by exploiting GPS, embedded sensors, and wireless local area network (WLAN) technologies for positioning, data collection, and communication. However, for large-scale tracking wireless sensor networks (WSN) that track the random movement of people, the continuous usage of a person smartphone's GPS and Wi-Fi is not an energy-efficient solution. In order to save energy and prolong the lifetime of the network, nearby mobile smartphones are traditionally grouped to form Bluetooth small groups (clusters) classified into a single level of Bluetooth clustering [1, 2]. In Bluetooth specifications, a cluster, also called a piconet, is a small network that consists of one master (cluster head) and up to seven slaves (cluster members). Cluster members communicate locally with their cluster head via low-energy Bluetooth, while only the cluster head uses energy-consuming Wi-Fi for long-range communications with the server [3].

In this paper, an energy-efficient solution is proposed for large-scale tracking systems without continuously using everybody's smartphone's GPS and Wi-Fi. The proposed approach is used to further reduce energy consumption in large-scale Bluetooth networks by building hierarchical clusters to further reduce the number of masters that use Wi-Fi communications. This is equivalent to extending the number of cluster members per master, and it is done by creating a two-level hierarchical clustering structure, which is called a "Scatternet". A master node in a first-level cluster can be a slave in a second-level cluster, thus linking the two levels together. In the second level of the clustering hierarchy, the second-level master, which is called a super master, is responsible for providing positioning information as well as sharing data with all the members of its group and also with the back-end server. First-level members and masters use low-energy Bluetooth, while only second-level masters (super masters) use Wi-Fi. Since the super masters constitute only a small subset of the masters, the use of energy-consuming Wi-Fi is significantly reduced.

The proposed clustering technique is designed to achieve the following objectives: (1) higher positioning accuracy, (2) lower energy consumption and hence longer network lifetime, (3) reduced transmission delay, and (4) lower signal interference. This paper has several main contributions. The first main contribution is formulating a new mathematical programming model to minimize mutual interference and energy consumption in large-scale mobile tracking networks based on optimum bi-level clustering hierarchy. The second contribution is developing an efficient heuristic method, called the hierarchical clustering algorithm, to solve the problem by constructing a two-level clustering hierarchy. The third contribution of this paper is constructing a new simulation model to measure interference between different Bluetooth signals, in addition to interference between Bluetooth and Wi-Fi signals.

The remainder of the paper is organized as follows. In section 2, recent literature related to mobile tracking systems is reviewed. In section 3, the mathematical programming model of the problem is presented. In section 4, the proposed bi-level hierarchical clustering algorithm is described. In Section 5, the simulation model is described and its results are analyzed. In Section 6, the paper is summarized and suggestions for future research are made.

## 2. Literature Review

This section summarizes relevant previous work related to optimizing energy for large-scale tracking systems, focusing on clustering approaches for mobile Bluetooth networks. Many researchers approached the problem by using either Bluetooth or Wi-Fi as the short-range radio interfaces. However, none of these solutions are fully suitable for large-scale tracking applications in open areas with huge crowds, high mobility, and signal interference.

Clustering approaches have been frequently used in Wireless Sensor Networks (WSN). Abbasi and Younis [4] surveyed different clustering approaches for WSN and compared them based on several metrics such as convergence rate, stability, overlapping, energy efficiency, failure recovery, balanced clustering, and node mobility. However, Bluetooth technology was not considered in classifying the clustering schemes. Bandyopadhyay and Coyle [5] proposed an algorithm that minimizes the total energy consumed in the system by organizing sensors into a hierarchy of clusters that communicate together to transmit data to the data processing center. They implemented their approach assuming that the communication environment is contention-free and error-free, which made the algorithm fit for networks with a large number of nodes. Yoo and Park [6] presented a distributed clustering approach, called the Cooperative Networking protocol (CONET) for improving the energy efficiency of wireless networks. CONET dynamically clusters the nodes in the network according to the bandwidth, energy, and application type of each node. The clustering approach of Yoo and Park [6] is based on one-level clustering, assuming no mobility and no mutual interference.

Teng et al. [7] proposed an energy-saving collaborative method for solving multi-target tracking problems in WSN. Distributed cluster-based variational target tracking is used when the targets are far apart, and data association is used when the targets are close to each other. The method is designed to track a varying number of targets, by allowing the arrival of new targets and the departure of some existing targets. Leem et al. [8] proposed a mobile clustering procedure with three cost metrics: the residual battery energy, the physical data rate, and the number of members in the estimated cooperation group. The device which has highest cost metric value is chosen as a group header (GH) and the other devices act as group members (GMs). The GH receives the data from GMs via its Bluetooth interface and transmits the data to a WLAN access point (AP) by using its WLAN interface. This approach is not suitable for saving energy for a large-scale system, since it can lead to high interference when the number of nodes in the WLAN AP coverage is large.

Al-Kanj et al. [9] presented a comprehensive overview of energy-efficient peer-to-peer collaboration in mobile devices based on mobile clustering through wireless networks. They also presented an analytical study of the impact of factors such as the number (cluster size) of a set of mobile terminals that are interested in downloading the same content cooperatively. However, they did not consider finding the optimal cluster size to minimize energy consumption. Al-Kanj and Dawy [10] presented an analytical study of the impact of network parameters, such as the values of the sending and receiving energies per unit time and the cluster size of a set of cooperating mobile terminals, on long-range and short-range energy consumption. They considered a grid network, which is a frequently used model in ad-hoc and sensor networks, where it is sufficient to derive the best energy consumption expressions as a function of the network parameters. The approach does not consider the optimal size of mobile clusters to minimize energy consumption.

Baranidharan and Santhi [11] used a fuzzy clustering approach to reduce energy consumption in WSN by balancing the load among the cluster heads. Distributed Unequal Clustering using Fuzzy logic (DUCF) is used to select cluster heads and assign them to clusters of different sizes. Based on the residual energy, node degree, and distance to base station, DUCF selects the cluster heads and determines the number of members in each cluster. Oren et al. [12] proposed the Adaptive Distributed Hierarchical Sensing (ADHS) algorithm. To optimize energy consumption, ADHS uses a proximity-traversing-based algorithm, called Hierarchical Control Clustering (HCC). HCC consists of two main sub-processes, the first is the Tree Discovery process and the second is the Cluster Formation process. Aiming to balance network lifetime and coverage in WSN, Xu et al. [13] pursued three optimization objectives: minimum energy consumption, maximum coverage rate, and maximum equity of energy consumption. Two multi-objective evolutionary algorithms (MOEAs) were developed to solve the optimization model, one based on decomposition and the other based on genetic algorithms.

In addition to clustering, a variety of other techniques have also been used for optimization applications in WSN. Oh et al. [14] proposed a bandwidth aggregation (BAG) technique to design energy-efficient heterogeneous

Wi-Fi access networks. Their algorithm can reduce power consumption in mobile devices with Wi-Fi access networks, but it does not use Bluetooth. Farrag et al. [15] developed a new localization technique for WSN by integrating the received signal strength indicator (RSSI) method with the social network analysis (SNA) method. Since the RSSI method is limited to pairs of nodes within communication range, it is supplemented by SNA to improve localization accuracy.

According to the above literature review, previous published works have not sufficiently considered some realistic aspects of co-existing multiple Bluetooth clusters with mobility and signal interference. In this paper, a new approach will be presented for large-scale systems in environments where communication is affected by interference and errors. The new approach for designing energy-efficient wireless mobile tracking systems is based on bi-level hierarchical clustering, and it has several realistic features that have not been addressed simultaneously before. Two levels of mobile clusters are developed, in which Wi-Fi communication is limited to second-level masters (super masters) that constitute a very small subset of the total number of nodes. In addition, the new approach considers both Bluetooth and Wi-Fi signals, and optimizes the number of clusters and the cluster formation (i.e. assignment of a master and members for each cluster). Moreover, the method considers the interference between Bluetooth and Wi-Fi signals as well as between different Bluetooth signals for multiple nodes co-existing in the same area.

## 3. The Mathematical Programming Model

This section presents a mathematical programming model to optimize the performance of large-scale tracking systems based on bi-level hierarchical clustering structure. In order to achieve the goals of efficient energy consumption and good quality of service, the following objectives must be optimized:

(1) Minimizing the number of clusters (masters) for the first-level of the clustering hierarchy.

(2) Minimizing the total distance between masters and slaves for the first-level of the clustering hierarchy.

(3) Minimizing the number of clusters (super masters) for the second-level of the clustering hierarchy.

(4) Minimizing the total distance between super masters and masters for the second-level of the clustering hierarchy.

The first objective, minimizing the number of the clusters, leads to reducing the energy consumption, which in turn leads to maximizing the network lifetime. Moreover, minimizing the number of clusters reduces channel access congestion, which reduces the interference among Bluetooth-based clusters as well as Bluetooth/Wi-Fi communications when they are employed in the same area. The second objective, which is minimizing the total distance between first-level masters and slaves, leads to higher accuracy of positioning. The master node represents the location information of all its slaves. Therefore, communicating via short-range radio interfaces such as Bluetooth is more accurate for reporting locations than communicating via long-range radio interfaces. The expected error in positioning is ±10m, which is the maximum range of the Bluetooth signals. Furthermore, shorter distances reduce the energy consumption and the transmission delay of Bluetooth networks, since communicating via short-range radio interfaces such as Bluetooth consumes lower power than communicating via long-range radio interfaces. The third objective is similar to the first objective, and it has similar benefits, but it applies to the second level of the network clustering hierarchy. Likewise, the fourth objective is similar to the second objective and has similar advantages, but it applies to the second level of the hierarchy.

### 3.1. Definition of Symbols

Let $i = 1$ to $N$ denote the slave number, $j = 1$ to $N$ denote the master number, and $C_{ij}$ denote the distance between node $i$ and node $j$. Let $WF_j$ denote the availability of Wi-Fi service in node (user's smartphone) $j$ as defined in (1). The user's battery level ($BL_j$) is defined as in (2). Equations (3-6) define the decision variables of the optimization model. Equations (3) and (4) describe the decision variables for the first level of the network clustering hierarchy, while equations (5) and (6) describe the decision variables for the second level.

$$WF_j = \begin{cases} 1, & \text{if device } j \text{ has Wi} - \text{Fi} \\ 0, & \text{otherwise} \end{cases} \tag{1}$$

$$BL_j = \begin{cases} 1, & \text{if battery level of device } j \text{ is} \geq 50\% \\ 0, & \text{otherwise} \end{cases} \tag{2}$$

$$X_{ij} = \begin{cases} 1, & \text{if device } i \text{ is a slave to master } j \\ 0, & \text{otherwise} \end{cases} \tag{3}$$

$$Y_j = \begin{cases} 1, & \text{if device } j \text{ is assgined as a master.} \\ 0, & \text{otherwise} \end{cases} \tag{4}$$

$$V_{ij} = \begin{cases} 1, & \text{if master } i \text{ is a slave to super master } j \\ 0, & \text{otherwise} \end{cases} \tag{5}$$

$$W_j = \begin{cases} 1, & \text{if master } j \text{ is assgined as a super master} \\ 0, & \text{otherwise} \end{cases} \tag{6}$$

### 3.2. The Optimization Model

The complete mathematical model for this problem is given by the set of equations (7)-(15). The objective function (7) is to minimize Z, which consists of four terms. The first two terms aim to minimize the number of clusters (masters) and the total distance between masters and slaves for the first level of the network hierarchy. The third and fourth terms aim to minimize the number of clusters (super masters) and the total distance between super masters and masters for the second level of the hierarchy.

The objective function (7) is to be optimized subject to nine constraints. The first constraints (8) ensure that every slave has a master. Constraints (9) and (10) limit first-level cluster size to 8 (1 master plus 7 slaves) and ensure that first-level cluster members are within the Bluetooth range of 10m, respectively. Constraints (11) ensure that every master has a super master. Constraints (12) and (13) limit second-level cluster size to 8 (1 super master and 7 first-level masters) and ensure that second-level cluster members are within the Bluetooth range, respectively. Constraints (14) ensure that each super master must be already a master. The last two constraints, (15) and (16), ensure that the master has Wi-Fi connection and a battery level greater than or equal to 50%, respectively. The fixed cost of each master and super master is denoted by $F$, and it is set equal to 100.

$$\text{Minimize } Z = \sum_{i=1}^{N}\sum_{j=1}^{N} C_{ij}X_{ij} + \sum_{j=1}^{N} F_j Y_j + \sum_{i=1}^{N}\sum_{j=1}^{N} C_{ij}V_{ij} + \sum_{j=1}^{N} F_j W_j \tag{7}$$

Subject to

$$\sum_{j=1}^{N} X_{ij} = 1, i = 1 \dots N \tag{8}$$

$$\sum_{i=1}^{N} X_{ij} \leq 8 Y_j, j = 1 \dots N \tag{9}$$

$$\sum_{j=1}^{N} C_{ij}X_{ij} \leq 10, i = 1 \dots N \tag{10}$$

$$\sum_{j=1}^{N} V_{ij} \leq Y_i, i = 1 \dots N \tag{11}$$

$$\sum_{i=1}^{N} V_{ij} \leq 8W_j, j = 1 \dots N \tag{12}$$

$$\sum_{j=1}^{N} C_{ij}V_{ij} \leq 10Y_j, i = 1 \dots N \tag{13}$$

$$W_j \leq Y_j, \quad j = 1 \dots N \tag{14}$$

$$Y_j \leq WF, \quad j = 1 \dots N \tag{15}$$

$$Y_j \leq BL, \quad j = 1 \dots N \tag{16}$$

## 4. The Hierarchical Clustering Approach

In very large-scale mobile tracking applications, the optimum solution of the mathematical model presented above can be time-consuming. Therefore, a faster heuristic approach for hierarchical clustering is presented here for very large-scale networks. The running times of the heuristic and the optimum solutions are compared in section 5. It is assumed to have a large set of moving nodes (individuals, each with their own cell phone). This set of nodes will be divided into small clusters (groups) of nearby smartphones arranged in two hierarchical levels. To save energy, only second-level masters (super masters) use Wi-Fi to share positioning information and health data with the back-end server. Each cluster resides within a small area, so communication within each (first-level or second-level) cluster is done via a short-range, low-energy radio interface, such as Bluetooth.

Fig. 1 illustrates the concept of the two-level hierarchical clustering algorithm. To preserve energy, low-energy short-range Bluetooth signals via personal area networks (PAN) are used for communication between masters and slaves and between masters and super-masters. In this paper, the analysis is based on using Bluetooth version 4.2, also known as Bluetooth Low Energy (BLE) or Bluetooth Smart. This version has several useful features such as higher speed, greater capacity, and enhanced security. Version 4.2 allows only trusted owners to track device locations and confidently pair devices. It also facilitates Internet of Things (IoT) applications due to its low power consumption and efficiency in transmitting data over the Internet [3]. On the other hand, high-energy, long-range Wi-Fi signals via WLAN are used only for communications between the super-masters and the back-end server.

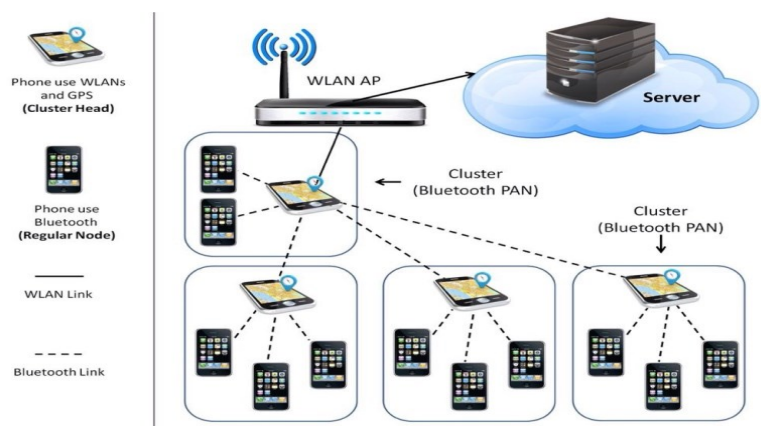

Fig. 1. Proposed solution (Hierarchical Clustering Bluetooth Network).

*4.1. Hierarchical Clustering Algorithm*

A heuristic clustering algorithm is developed to optimize energy and reduce interference for large-scale Bluetooth networks based on bi-level hierarchical mobile clustering. The algorithm, described below, aims to satisfy the requirements for tracking purposes in large-scale environments. Therefore, the proposed bi-level clustering algorithm focuses specifically on node battery level and Wi-Fi connection availability. In order to achieve the best performance, the following requirements have to be met:

- Each node makes its own decisions based on its local information.
- Each node can be either a cluster member that belongs to exactly one cluster (slave for first-level master), a first-level master (slave for the super master), or a second-level master (super master).
- Clusters must include all nodes, without any overlap (common nodes) between different clusters.
- Message exchange should be efficient in order to meet clustering processing requirements.

The heuristic Hierarchical Clustering Algorithm iteratively constructs a bi-level clustering structure for all nodes in the network. When all nodes are booted up, each node will broadcast its battery level and Wi-Fi connection availability to all nodes within its range. First, in the construction of first-level clusters, all nodes that have Wi-Fi connection availability are eligible to be cluster heads (masters) or cluster members (slaves), while the nodes that do not have Wi-Fi are only eligible to be cluster members. Next, among the nodes with Wi-Fi connection, the node that has the highest battery level will be chosen as the first-level master of the particular cluster, and the closest (up to seven) nodes can join as slaves. The construction process continues until each node is assigned as either a master or a slave that belongs to exactly one cluster at the end of the first-level clustering procedure.

After that, the construction of second-level clusters begins. The first-level master which has the highest battery level will be chosen as a super master (second-level cluster head) of the first-level masters, and the closest (up to seven) first-level masters can join as second-level cluster members (slaves for the super master). The construction process continues until each node is assigned either as a slave, a master, or a super master. Slaves and masters use low-energy Bluetooth technology, while only super masters use Wi-Fi technology. The pseudo code of the Hierarchical Clustering Algorithm, which constructs a two-level hierarchical clustering structure for mobile networks, is shown Table 1.

Table 1: Pseudo code of the algorithm

| | | | |
|---|---|---|---|
| 1: | **If** $N > 0$ ($N$: Number of nodes) | 23: | ----***Start second-level clustering***---- |
| 2: |   **for** $i = 1: N$ // **first-level clustering** | 24: |   **for** $i = 1: K$ ($K$: Number of first-level Masters) |
| 3: |     **if** ($i$ has Wi-Fi) | 25: |     **if** ($i$ has highest battery level $BL$) |
| 4: |       Update possible Master | 26: |       Update become super master |
| 5: |     **else** | 27: |     **else** |
| 6: |       Update possible Slave | 28: |       Update possible second-level member |
| 7: |   **end if** | 29: |   **end if** |
| 8: |   **if** ($i$ has highest battery level $BL$) | 30: |   **for** $j = 1$: possible second-level member |
| 9: |     Update become Master | 31: |     **if** ($C_{ij} \leq 10$)// Bluetooth range |
| 10: |   **end if** | 32: |       Update $M$-possible second-level member |
| 11: |   **for** $j = 1$: possible Slave | 33: |   **end if** |
| 12: |     **if** ($C_{ij} \leq 10$)// Bluetooth range | 34: |   **if** ($M \leq 7$) |
| 13: |       Update $M$-possible Slaves | 35: |     Update all $M$ become second-level members |
| 14: |   **end if** | 36: |     Cluster size = $M$ |
| 15: |   **if** ($M \leq 7$) | 37: |   **else** |
| 16: |     Update all $M$ become Slaves | 38: |     Update closest 7 become second-level members |
| 17: |     Cluster size = $M$ | 39: |     Cluster size = 7 |
| 18: |   **else** | 40: |   **end if** |
| 19: |     Update closest 7 become Slaves | 41: |   Update $K = K$ - Cluster size |
| 20: |     Cluster size = 7 | 42: | **end if** |
| 21: |   **end if** | | |
| 22: |   Update $N = N$ - Cluster size | | |

## 4.2. Time Scheduling for Efficient Data Transmission

After clusters are formed and masters and super masters are selected, the process of data exchange starts. This process takes place between slaves and first-level masters, first-level masters and super masters, and super masters and the back-end server. The super masters use Wi-Fi to transmit data to the server, where this data is processed and stored. Then, after a given period, new clusters are formed at two levels and new masters and super masters are selected. The order of the list will be refreshed periodically and the two-level clustering hierarchy will be periodically reconstructed. This procedure guarantees fair load distribution among multiple devices, attains maximum throughput and lifetime of the network, and avoids draining the batteries of a few super master devices.

The Bluetooth technology provides two types of data transmissions between devices in a piconet: synchronous connection oriented (SCO) and asynchronous connection less (ACL). SCO is point-to-point transmission data between the master and a single slave. The master reserves time slots to ensure that the capacity is available for an SCO link, while ACL connection can use any time slot. ACL can also be used as a point-to-point connection, but the master can broadcast the data to multiple slaves and the slave can only send data when requested to do so by the master. In the proposed approach, ACL transmissions are used within each piconet to communicate between the master and all the slaves. All of the devices connected to a piconet use the same frequency-hopping schedule and are controlled by the master.

The channel is divided into 625 microsecond time slots and it is shared between the master and the slaves of each cluster using the simple rule that the master transmits in even time slots and the slaves transmit in odd time slots [3]. Therefore, two time slots (1250 μs) are needed for each slave to transmit its data to the master. It is important to avoid interference from other Bluetooth devices using the industrial, scientific, and medical (ISM) radio band. For that reason, a simple time schedule is used for each device in a second-level Bluetooth cluster. According to this schedule, time is divided between two modes of operation: active mode and sleep mode. The time for each mode is 1250 microseconds, and each device can send its data only when it is in the active mode.

An example of time scheduling for first-level clusters of the network is depicted in Fig. 2. The first-level of the network hierarchy consists of three clusters (piconets), and each cluster consists of one master and seven slaves. The three masters are responsible for transmitting data via Wi-Fi for processing to the back-end server. It can be observed from Fig. 2 that the delay time remains constant as the number of clusters increases. However, the quality of signal transmissions decreases as the number of clusters increases due to mutual interference among nearby clusters. When two or more clusters send Bluetooth signals at the same frequency, the signals may collide and become distorted or lost. Communication quality is measured by the frame error rate (FER), which is the proportion of wrong or lost data out of the total data received. Obviously, FER increases as the number of clusters increases, leading to loss of some data that may be critical, especially for tracking purposes.

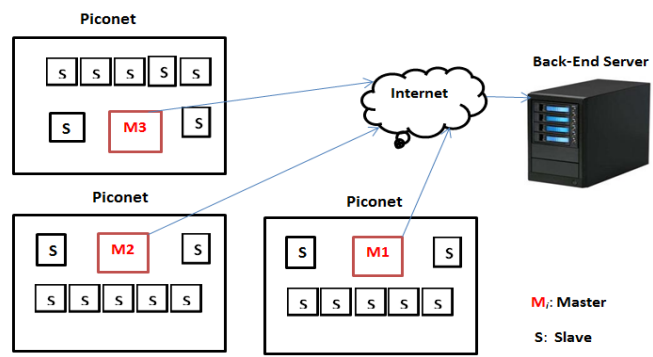

Fig. 2. Example of time scheduling for first-level clusters of the network hierarchy.

An illustration of an example of the time scheduling process for second-level clusters in the Bluetooth network hierarchy is shown in Fig. 3. In this figure, the second-level of the hierarchy consists of three clusters, where each cluster consists of one master and seven slaves. Only the super master is responsible for transmitting data via Wi-Fi for processing to the back-end server. Fig. 3 shows that the delay time increases as the number of clusters increases. Adding a second level to the network hierarchy reduces the number of masters that communicate directly with the back-end server, which decreases the frame error rate (FER). Therefore, the proposed two-level hierarchical clustering is more efficient than one-level clustering, especially for highly populated areas.

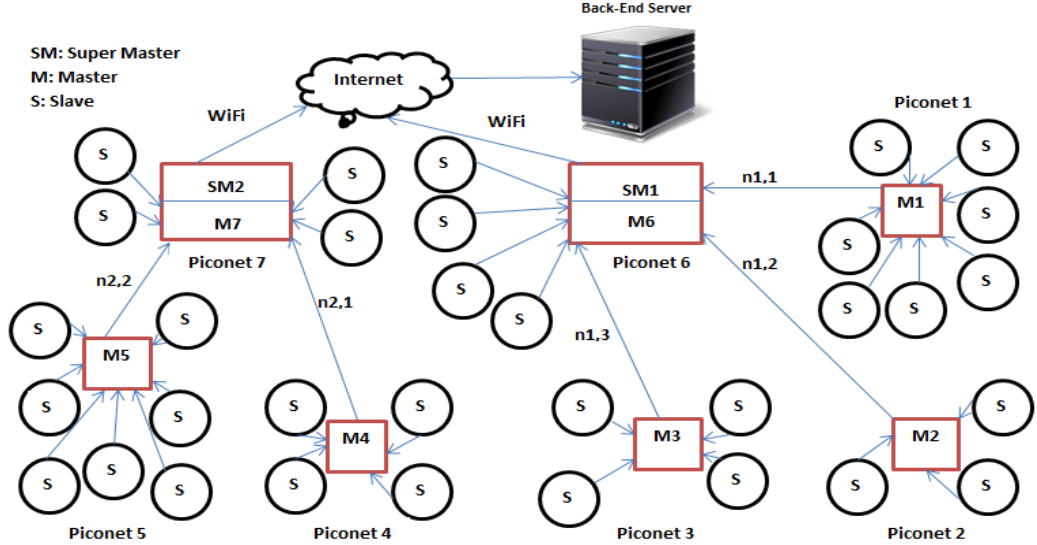

Fig. 3. Example of time scheduling for second-level clusters of the network hierarchy.

*4.3. Algorithm Analysis*

*Lemma-I*:

*For first-level clusters using Bluetooth as a medium among themselves and Wi-Fi as a medium to communicate with a server, the maximum delay will not exceed $D_{max}^1$.*

*Proof:*

Assume $N$ nodes forming $M$ first-level clusters using Bluetooth to communicate among themselves, with $M$ master nodes using Wi-Fi to communicate with a server. Denote $n_i^1$ as the number of nodes in the $i^{th}$ first-level cluster, then $TTS_i^1$ which is the time cluster $i$ needs to transmit data to the server can be computed as in Eq. (17).

$$TTS_i^1 = n_i^1 T \qquad (17)$$

Here, $T$ denotes the standard Bluetooth cycle duration, i.e. $T = 1250$ µs (Bluetooth SIG, 2017). Once this time passes, the master node can send the collected data including its own data along with its position to the server. Since the size of clusters differs, the maximum delay that a node may incur is computed as follows.

$$D_{max}^1 = \max\{n_i^1\}T, i = 1,2,\dots,M \qquad (18)$$

$$TD^1 \leq D_{max}^1 \qquad (19)$$

$D_{max}^1$ is the maximum delay for the first-level clusters in the network hierarchy, which is dominated by the cluster that had the maximum number of nodes among all first-level clusters ($\max\{n_i^1\}$). $TD^1$ is the total delay of all clusters in the first-level of the network hierarchy, which is equal to $D_{max}^1$.

Example-I:

For the one-level clustering system shown in Fig. 2, there are 3 clusters, each containing 7 members and a master. The maximum number of nodes in each cluster is $\max\{n_i^1\} \leq 7$. Therefore, the total delay of all clusters $TD$ is equal to the maximum delay $D_{max}^1 = 7T$.

*Lemma-II*:

*For second-level clusters using Bluetooth as a medium among themselves and Wi-Fi as a medium to communicate with a server, the maximum delay will not exceed $D_{max}^2 = \max\{TTS_i^2\}$.*

*Proof:*

Assume $N$ nodes forming $M^1$ first-level clusters and $M^2$ first-level clusters using Bluetooth as a medium among themselves, with $M^2$ super master nodes using Wi-Fi as a medium to communicate with a server. Denote $n_{ij}$ as the number of nodes in the $i^{th}$ first-level cluster belonging to the $j^{th}$ second-level cluster, then $TTS_i^2$, which is the time the second-level cluster $i$ needs to transmit data to the server can be computed as in Eq. (20).

$$TTS_i^2 = \left[\left(\sum_{j=1}^{M^1} n_{ij} + 1\right) - 1\right]T, i = 1,2,\dots,M^2 \qquad (20)$$

In (20), the constant 1 is added to each $n_{ij}$ to include the data of all first-level master nodes. Since the second-level master (super master) node is also a first-level master, the constant 1 is subtracted from the total delay. Furthermore, each super master node also has to schedule the transmissions of its members. Hence, the total delay ($TD^2$) for transmitting the complete collected data to the server excluding Wi-Fi transmission can be computed as follows.

$$TD^2 \leq \max\{TTS_i^2\} \qquad (21)$$

Example-II:

For the two-level clustering system shown in Fig. 3, there are 7 first-level clusters and 2 second-level clusters. Therefore:

$$TTS_1^2 = [(7+1) + (3+1) + (4+1) + 4]T = 21T$$

$$TTS_2^2 = [(4+1) + (7+1) + 4]T = 17T$$

Since every super master node is independent of other super master nodes, each one will send once its data are ready. Hence, the maximum delay for this example is:

$$TD^2 \leq \max\{TTS_1^2, TTS_2^2\} = 21T$$

## 5. Performance Evaluation

In this section, the performance of the proposed hierarchical clustering approach is evaluated by comparing to the optimal solution of model (7)-(16) and by using the simulation model. First, the optimal solution is presented and then the simulation results are discussed.

### *5.1. Optimum Solution*

The above formulation was solved under three different scenarios using version 24.3.3 of GAMS (General Algebraic Modeling System) [16]. The first scenario tackles the problem by considering only the first two terms in the objective function that aim to minimize the number of clusters (masters) and the total distance between masters and slaves for the first level of the clustering hierarchy. The second scenario considers all four terms in the objective function that aims to minimize the number of clusters and the total distance between masters and slaves for both levels of the hierarchy. The third scenario considers all four terms of the objective function, and also applies sensitivity analysis by considering two different values for the number of nodes, namely 700 and 800. This is done by changing the second-level cluster size, i.e. the right-hand side (RHS) of constraints (12).

All above-described scenarios have been analyzed under the following setup. The optimal value of the objective function is calculated by GAMS MIP solver, assuming the following different values for the number of nodes: $N$ = 100, 200, 300, 400, 500, 600, 700, and 800.

Fig. 4 shows the optimum number of clusters for both scenario 1 and scenario 2. For scenario 1, in which only first-level clustering objectives are considered, Fig. 4 shows the number of clusters (masters) for the first level of the network clustering hierarchy as a function of the number of nodes $N$. It can be observed that for 100 nodes, only 14 masters are needed. However, for 800 nodes, 100 masters are required. The maximum number of nodes per cluster is 800/100 = 8, implying up to 7 slaves per master, which is acceptable for a large-scale system.

Fig. 4 also shows the results for scenario 2, in which the objectives of both levels of the clustering hierarchy are considered. Fig. 4 displays the number of clusters (super masters) for the second level of the network clustering hierarchy. As expected, the number of the clusters in the second level is much smaller than in the first level. For example, for 100 nodes, only 2 clusters (super masters) are required in the second level of the hierarchy, compared to 14 clusters in the first level. When there are 800 nodes, only 13 clusters are needed in the second level compared to 100 clusters in the first level. Therefore, using second-level (super) masters to communicate with the back-end server via Wi-Fi leads to significant savings in energy compared to communicating through first-level masters.

Fig. 5 presents the results for scenario 3, showing sensitivity analysis in terms of the number of nodes. Two values for the total number of nodes in the system are considered: 700 and 800. Fig. 5 shows the optimal number of clusters versus the cluster size of the second level of the network hierarchy. The horizontal axis in the figure represents the constant in the RHS of constraint (12), which is varied from 1 to 8. For 700 nodes, the number of clusters is minimum when the cluster size is equal to 8, corresponding to 11 clusters. For the case of 800 nodes, the number of clusters is also minimum when the cluster size is equal to 8, but it corresponds to 13 clusters. The small number of second-level clusters shows that the two-level hierarchical clustering approach is effective for tracking applications, especially in highly populated areas.

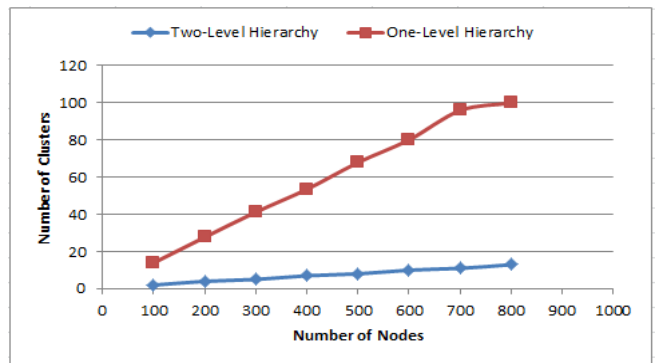

Fig. 4. Optimal number of clusters for scenario 1 and scenario 2.

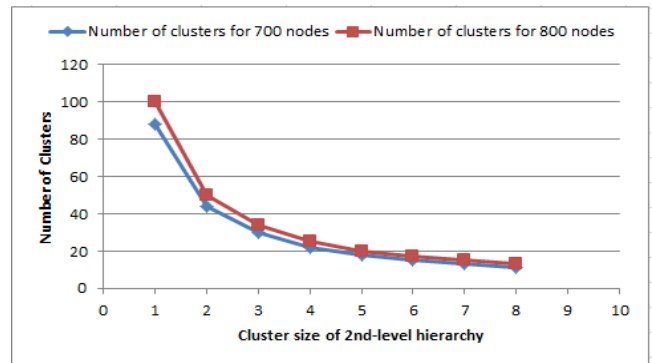

Fig. 5. Optimal number of clusters for scenario 3.

*5.2. Simulation Experimental Setup*

In this section, the performance of the proposed bi-level clustering heuristic approach is evaluated by using a MATLAB Simulink simulation model. This tool is commonly used to analyze models of radio frequency mechanisms of Bluetooth transceivers, as done by Golmie et al. [17] and by Song et al. [18]. Each transceiver consists of a binary data generator, a Gaussian frequency shift keying (GFSK), a pseudo-random number generator to create frequency hopping, and a matching receiver [19]. To introduce noise, the 802.11 packet block is generated as an interference source.

In the simulation experiments, the Bluetooth full duplex voice and data transmission model is used, since it is more realistic than the standard Bluetooth model. Fig. 6 depicts the full duplex communication process between two Bluetooth devices. The model consists of a sender and a receiver, where one of them should be assigned as the master and the other one as the slave. Furthermore, an 802.11b packet block is created as an interference source by a separate independent block. Both data packets and voice packets can be transmitted between the two devices. The supported voice packet types are: HV1, HV2, HV3 and SCORT, and the supported data packet type is DM1 [20]. Fig. 6 shows the configuration of the simulation model used for performance evaluation of the Bluetooth network. Model components include: scatternet, piconet, interference source from 802.11, and modules for measuring important performance parameters of the system.

Fig. 7 illustrates the new interference model that consists of two clusters existing in the same area. The signal Tx11 denotes the transmitting power signal to cluster 1 and the signal Tx21 denotes the signal to cluster 2. These two signals are used to find the effect of interference among different Bluetooth signals. In addition, the signal Tx_802 denotes the 802.11b packet block used as an interference source to find the effect of the interference between Bluetooth and Wi-Fi signals [21]. By adding the transmitting power signal to each device, it is possible estimate the effect of interference among different signals. This is done by calculating the free space path loss as a function of the distance to the surrounding clusters, which varies randomly from 0.1m to 10m.

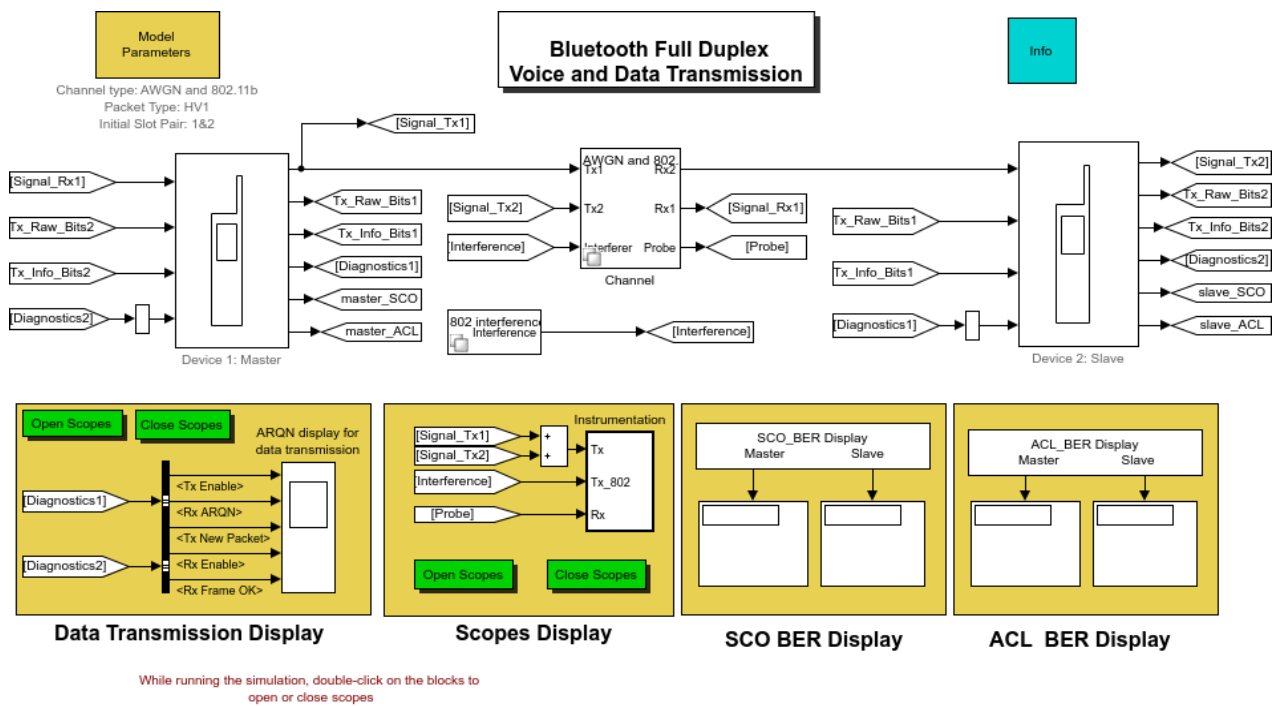

Fig. 6. Bluetooth Full Duplex Voice and Data Transmission model.

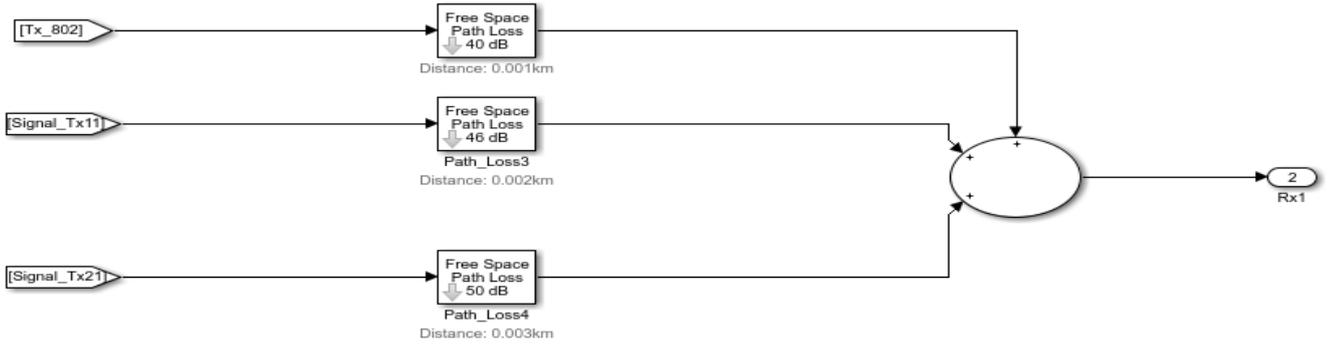

Fig. 7. Adding the transmitting power signal to the model.

In order to study Bluetooth piconets (clusters) for highly populated areas, the simulation model considers $P$ Bluetooth piconets existing together in an area of size $10 \times 10$ m$^2$. Therefore, each piconet experiences possible interference by $(P – 1)$ other piconets. If two or more piconets send out a packet (an information message) on the same frequency band at any time, then the corresponding packets collide and are considered lost. As per Bluetooth standards, all clusters employ a frequency-hopping technique in which a random channel is selected from 79 possible frequency channels. The simulation model is able to capture the interference between different Bluetooth packets and between Bluetooth and Wi-Fi packets when they use the same frequency range [21].

Fig. 8 displays the average frame error rate (FER) for one piconet, consisting of one master and one slave, which is affected by $P – 1$ piconets. It is assumed that the flow data volume of each Bluetooth device is fixed in the piconet. Using the DM1 data packet type, the average Frame Error Rate (FER) of the piconet is calculated for each master and slave. Changing the distance to the surrounding piconets randomly (from 0.1 to 10 meters), the average FER is calculated for 20 runs of the simulation model. The aim of repeating each simulation experiment 20 times using different random distances and then calculating the average is to obtain a 95% confidence interval. Fig. 8 shows that the average FER of the master is greater than the average FER of the slave. It also shows, as expected, that the average FER increases when the number of the clusters increases and causes higher interference. Therefore, the model is designed to minimize the number of clusters in order to reduce the channel access congestion and hence reduce interference among different signals.

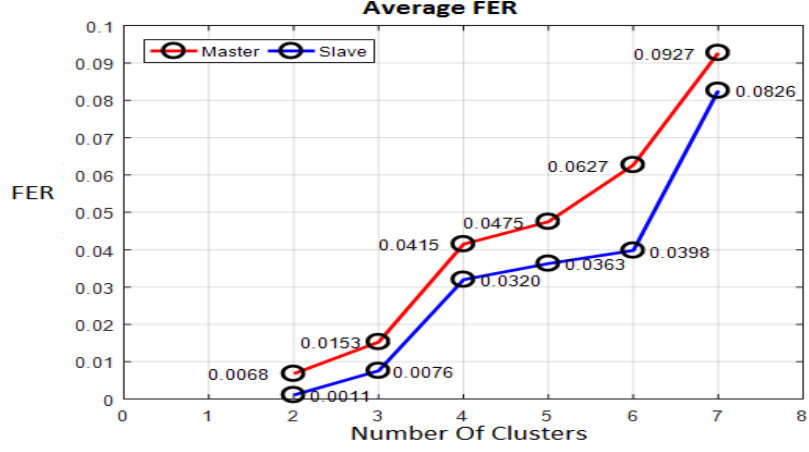

Fig. 8. Average frame error rate of for multiple Bluetooth coexisting clusters.

*5.3. Performance Metrics*

Performance was compared for three methods: the direct approach, the two-level hierarchical clustering algorithm, and the optimal GAMS solution. In the direct approach, there is no clustering as each node has Wi-Fi and GPS connections, and it transmits its own data directly to the back-end server. Relative performance of the three methods was evaluated using Matlab. It is assumed that each node can send data traffic at a rate of 1,000 kbps and it can send frames with sizes up to 20 bytes. For the hardware (Bluetooth/Wi-Fi) energy consumption parameters, the values specified by Yoo and Park (2011) were used. In order to achieve 95% confidence interval, each simulation experiment was repeated 10 times using different random topologies.

For each simulation run, the total energy consumption and throughput were calculated as follows for different values of the number of nodes ($N$ = 100, 200… 800):

$TE = TE_{CH} + TE_{CM} + TE_{idle}$  (22)

$G = N \times L \times R(1 - FER)$  (23)

$EF = G/TE$  (24)

Where:

$TE$ is the total energy consumption of all nodes. $TE_{CH}$ is energy consumption by cluster heads. $TE_{CM}$ is energy consumption by cluster members. $TE_{idle}$ is energy consumption by idle nodes. $G$ is throughput, which is defined as the total number of successfully received bits. $R$ is the frame rate per second. $N$ is the total number of nodes. $L$ is the frame length (size). $FC$ is the frame correction rate, where ($FC = 1 - FER$); $EF$ is energy efficiency.

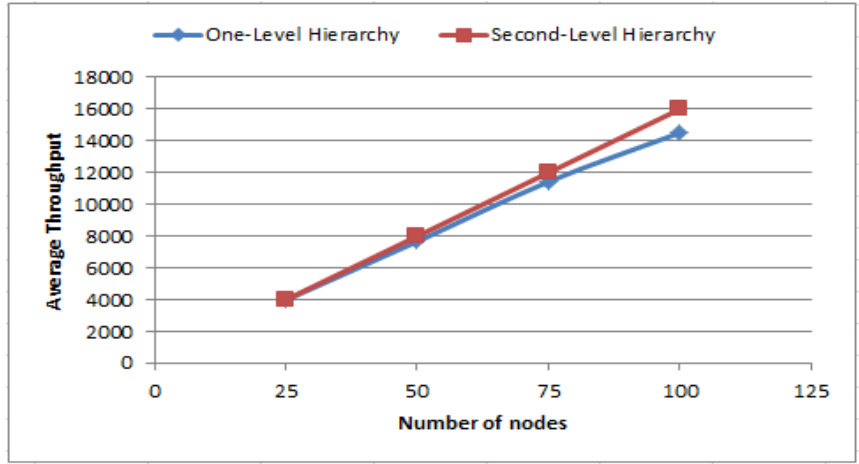

Fig. 9. Average throughput of the first-level and second-level clusters.

Fig. 9 shows the average throughput of both levels of the clustering hierarchy assuming different values of the total number of nodes. It is assumed that the frame size is equal to 20 bytes, which is sufficient to send health information messages. It can be observed from Fig. 9 that the throughput for the second clustering level is better than the throughput for the first level, especially as the number of nodes increases. This confirms the advantage of using second-level clusters, especially for highly populated areas.

*Lemma III:*

*The bi-level clustering effective throughput is higher than single-level clustering.*

*Proof:*

Assume $N$ nodes forming $M^1$ first-level clusters and $M^2$ first-level clusters using Bluetooth as a medium among themselves, with $M^2$ super master nodes using Wi-Fi as a medium to communicate with a server. Since each super master node sequentially schedules the transmission slots for its member, no signal collisions will occur. Therefore, the whole set of members belonging to the $j^{th}$ second-level cluster is effectively considered as one piconet in terms of mutual interference (i.e. no interference takes place among these members). Accordingly, the limiting factor in terms of interference is $M^2$ (the number of second-level clusters). Then, the total throughput in this case can be computed as follows. Denote $G^2(N)$ as the bi-level effective throughout,

$$G^2(N) = N{\times}L{\times}R[1 - FER(M^2)] \qquad (25)$$

On the other hand, for single-level clustering, the key factor is the number of first-level clusters ($M^1$)

existing in the same interference range. Therefore, the single-level effective throughout $G^1(N)$ can be computed by:

$$G^1(N) = N \times L \times R[1 - FER(M^1)] \quad (26)$$

But,

$$M^1 > M^2$$

So,

$$FER(M^1) > FER(M^2)$$

Therefore,

$$G^1(N) < G^2(N)$$

Example –III:

For the bi-level clustering system shown in Fig. 3, let $R$ = 1 frame/s, $N$ = 35, and $L$ = 20 bytes. We can observe from Fig. 3 that $M^1 = 7$ and $M^2 = 2$. From Fig. 8, FER (7) = 0.0927, while FER (2) = 0.0068. then:

$$G^1(K) = 35 * 20 * 8 * (1 - 0.0927) = 5081 \; bits$$

$$G^2(K) = 35 * 20 * 8 * (1 - 0.0068) = 5562 \; bits$$

The bi-level throughput is enhanced by 10%.

## *5.4. Simulation Results*

Fig. 10 and Table 2 show the average total energy consumption for the direct approach, the heuristic bi-level hierarchical clustering algorithm, and the optimal GAMS solution of the integer programming model. Fig. 10 shows that the total energy consumption of the two-level clustering solution is close to the minimum total consumption obtained by the optimal GAMS solution. In fact, the total energy of the hierarchical clustering algorithm becomes closer to optimality as the number of nodes increases. This is clear from Table 2, which shows a difference of 5% between the performance of the hierarchical clustering algorithm and the optimal GAMS solution when the number of nodes is equal to 100, but a difference of only 0.54% when number of nodes is equal to 800. This shows that the proposed two-level hierarchical clustering algorithm is capable of producing high-quality, near-optimum solutions for large-scale tracking problems.

As shown in Table 2, the energy consumption of the direct approach is 993.6% higher than the optimal consumption specified by GAMS when the number of nodes is equal to 100, and 1042.1% higher when the number of nodes is equal to 800. Clearly, the direct approach (without clustering) is not a practical solution method for large-scale tracking systems.

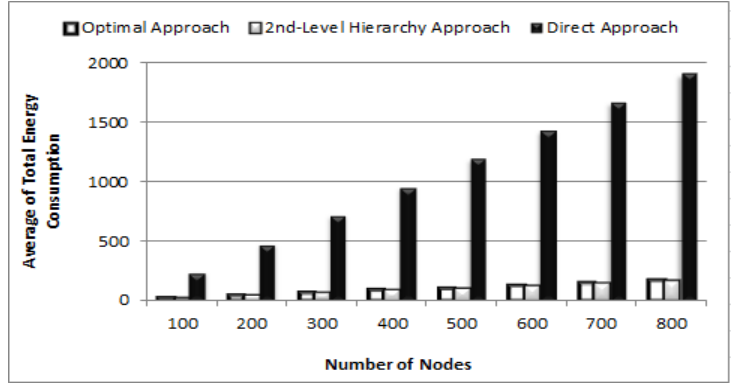

Fig. 10. Comparison of the average energy consumption for Scenario 2.

Table 2. Total energy consumption (Joules) of three solution methods for Scenario 2.

| #of nodes | Hierarchical Clustering Approach | Direct Approach | Optimal Approach GAMS | Comparison vs GAMS | |
|---|---|---|---|---|---|
| | | | | 2nd-Level Clustering Approach | Direct Approach |
| 100 | 22.9 | 238.41 | 21.8 | 5% | 993.60% |
| 200 | 45.6 | 476.82 | 43.5 | 4.80% | 996.10% |
| 300 | 65.6 | 715.23 | 62.9 | 4.29% | 1037.10% |
| 400 | 87.9 | 953.64 | 84.7 | 3.78% | 1025.90% |
| 500 | 107.3 | 1192.05 | 104.1 | 3.07% | 1045.10% |
| 600 | 128.8 | 1430.46 | 125.8 | 2.38% | 1037.10% |
| 700 | 147.5 | 1668.87 | 145.2 | 1.58% | 1049.40% |
| 800 | 167.9 | 1907.28 | 167 | 0.54% | 1042.10% |

Fig. 11 and Table 3 show average efficiency packet per Joule for the direct approach, the one-level clustering approach, and the two-level clustering approach. Fig. 12 shows that the efficiency packet per Joule of the two-level clustering approach is much better than the efficiency packet per Joule of the one-level clustering approach, especially when the number of nodes increases. This can also be seen from Table 3, which shows that the efficiency of two-level clustering is significantly higher than the efficiency of one-level clustering.

On the other hand, in the direct approach, the efficiency packet per Joule remains constant as the number of nodes increases. The reason is that in the direct approach, each node performs long-range data transmission to the back-end server. Therefore, the probability that each node has to transmit its data to the back-end server is the same regardless of the number of nodes. This is another reason to conclude that the direct approach is not suitable for handling high-data requirements of large-scale tracking systems.

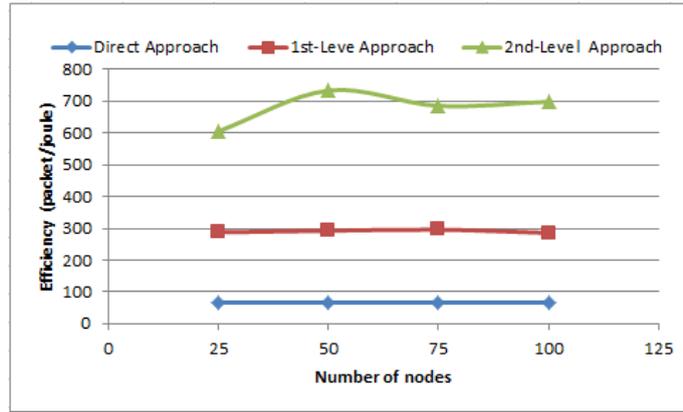

Fig. 11. Comparing the efficiency of direct, 1st and 2nd clustering approaches.

Table 3. Comparison of energy and efficiency of the three approaches

| Number of nodes | Throughput of (bps) | | | Energy Consumption of (Joule) | | | Efficiency of (packet/joule) | | |
|---|---|---|---|---|---|---|---|---|---|
| | Direct Approach | One-Level Approach | Two-Level Approach | Direct Approach | One-Level Approach | Two-Level Approach | Direct Approach | One-Level Approach | Two-Level Approach |
| 25 | 4000 | 3972.8 | 4000 | 59.6 | 13.8 | 6.6 | 67.1 | 287.9 | 606.1 |
| 50 | 8000 | 7668 | 8000 | 119.2 | 26.2 | 10.9 | 67.1 | 292.7 | 733.9 |
| 75 | 12000 | 11430 | 12000 | 178.8 | 38.6 | 17.5 | 67.1 | 296.1 | 685.7 |
| 100 | 16000 | 14516.8 | 16000 | 238.41 | 50.8 | 22.9 | 67.1 | 285.8 | 698.7 |

Fig. 12 shows the running times of bi-level clustering solutions obtained by the heuristic hierarchical clustering algorithm and the optimum integer programming model. The experiments were run on a laptop with an Intel® Core™ i5 2.50 GHz processor and a 4.0 GB RAM. Clearly, the heuristic algorithm is significantly faster, and hence it is able to handle much larger mobile tracking applications. This is an important advantage for the heuristic hierarchical clustering algorithm, in addition to its near-optimal performance shown in Table 2.

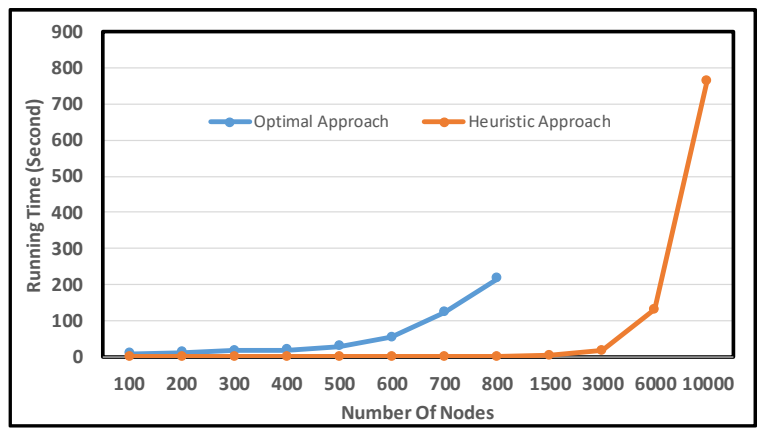

Fig. 12. Computation times for the optimal and the heuristic bi-level clustering solutions

## 6. Conclusions

This paper presented a new technique based on two-level hierarchal clustering for the optimization of energy consumption in Bluetooth mobile networks used in large-scale wireless tracking systems. An integer programming model is formulated to minimize the total distance between masters and slaves and the number of clusters for the first-level and the second-level of the bi-level clustering hierarchy. A hierarchical clustering heuristic algorithm is developed to efficiently generate near-optimal solutions. Simulation experiments show that both the integer programming model and the heuristic hierarchical clustering algorithm generate highly effective solutions, especially for large-scale tracking systems. These solutions improve the accuracy of positioning, decrease the energy consumption, and reduce the transmission delay. These solutions also maximize the network's lifetime and throughput, and decrease the levels of interference between different Bluetooth signals also between Bluetooth and Wi-Fi signals.

Based on the use of high-throughput and low-energy personal area networks (PAN), this work has optimized the energy efficiency for large-scale tracking networks. The new bi-level clustering approach can be used in tracking systems using smartphones, in which user locations and relevant pieces of information are reported to the server periodically. Future research can be directed to other areas still not tackled in this research such as optimizing with respect to some other objectives. In addition, there is a need to develop improved algorithms to design energy-efficient large-scale tracking systems, to further reduce the energy consumed by the processes of positioning and communication, and to guarantee a higher quality of service. Finally, to confirm practicality, empirical experiments needs to be carried out, in order to test the proposed method under various real-life conditions.

**Acknowledgments**

The authors Uthman Baroudi and Abdulrahman Abu Elkhail would like to acknowledge the support provided by the Deanship of Scientific Research (DSR) at King Fahd University of Petroleum and Minerals, under the grant RG1424-1.